\documentclass[12pt,preprint]{aastex}
\def\ltsima{$\; \buildrel < \over \sim \;$}
\def\lsim{\lower.5ex\hbox{\ltsima}}
\def\gtsima{$\; \buildrel > \over \sim \;$}
\def\gsim{\lower.5ex\hbox{\gtsima}}

\shorttitle{SNAP: Photometric redshifts}
\shortauthors{Dahlen et al.}

\begin{document}
\title{SuperNova Acceleration Probe (SNAP): Investigating 
                    Photometric Redshift Optimization}


\author{
Tomas Dahlen\altaffilmark{1}, 
Bahram Mobasher\altaffilmark{1},
Stephanie Jouvel\altaffilmark{2}, 
Jean-Paul Kneib\altaffilmark{2},
Olivier Ilbert\altaffilmark{2}, 
Stephane Arnouts\altaffilmark{2},
Gary Bernstein\altaffilmark{3},
and Jason Rhodes\altaffilmark{4,5}
}
\email{dahlen@stsci.edu}

\altaffiltext{1}{Space Telescope Science Institute, 3700 San Martin Drive, Baltimore, MD 21218.}
\altaffiltext{2}{Observatoire Astronomique de Marseille-Provence, F-13376 Marseille, France.}
\altaffiltext{3}{Department of Physics and Astronomy, University of Pennsylvania, Philadelphia, PA 19104.}
\altaffiltext{4}{California Institute of Technology, 1201E California Blvd., Pasadena, CA 91125.}
\altaffiltext{5}{Jet Propulsion Laboratory, 4800 Oak Grove Drive, Pasadena, CA 91109.}

\begin{abstract}
The aim of this paper is to investigate ways to optimize the accuracy of photometric redshifts for a SNAP like mission. We focus on how the accuracy of the photometric redshifts depends on the magnitude limit and signal-to-noise ratio, wave-length coverage, number of filters and their shapes and observed galaxy type. We use simulated galaxy catalogs constructed to reproduce observed galaxy luminosity functions from GOODS, and derive photometric redshifts using a template fitting method. By using a catalog that resembles real data, we can estimate the expected number density of galaxies for which photometric redshifts can be derived. We find that the accuracy of photometric redshifts is strongly dependent on the signal-to-noise (S/N) (i.e., S/N$>$10 is needed for accurate photometric redshifts). The accuracy of the photometric redshifts is also dependent on galaxy type, with smaller scatter for earlier type galaxies. Comparing results using different filter sets, we find that including the U-band is important for decreasing the fraction of outliers, i.e., ``catastrophic failures''. Using broad overlapping filters with resolution $\sim4$~gives better photometric redshifts compared to narrower filters (resolution $\gsim5$) with the same integration time. We find that filters with square response curves result in a slightly higher scatter, mainly due to a higher fraction of outliers at faint magnitudes. We also compare a 9-filter set to a 17-filter set, where we assume that the available exposure time per filter in the latter set is half that of the first set. We find that the 9-filter set gives more accurate redshifts for a larger number of objects and reaches higher redshift, while the 17-filter set is gives better results at bright magnitudes.  
\end{abstract}
 

\section{Introduction}
In recent years, there have been unprecedented progress in observational astronomy due, in large part, to the advent of large format and highly sensitive optical/infrared detectors. Installation of these cameras on 8m ground-based telescopes and space-borne facilities have enabled planning large and deep galaxy surveys, increasing the discovery space by over an order of magnitude. In particular, wide-area multi-waveband imaging from space, complemented by follow up ground-based observations, have provided extremely valuable datasets for studying diverse topics in observational astronomy and cosmology. For example, installation of the Advance Camera for Survey (ACS) on the Hubble Space Telescope (HST) has resulted in multi-waveband surveys of galaxies, including the Great Observatories Origins Deep Survey (GOODS - Giavalisco et al. 2004), COSMOS (Scoville et al. 2007) and the Hubble Ultra-Deep Field (HUDF- Beckwith et al. 2006). These surveys provide deep multi-waveband data covering large areas, used to study a number of issues concerning formation and evolution of galaxies, including: study of rest-frame properties of different populations of galaxies (e.g., Bundy et al. 2005; Grogin et al. 2005; Dahlen et al. 2007), search for the highest redshift (Kneib et al. 2004; Bouwens et al. 2005) and new population of galaxies (Wiklind et al. 2007), mapping of the dark matter distribution in strong lensing clusters (Smith et al. 2005; Broadhurst et al. 2005; Limousin et al. 2007; Richard et al. 2007), cosmological constraints from weak lensing (Massey et al. 2007a) and clustering of galaxies (McCracken et al. 2007) and the 3D large scale structure dark matter distribution (Massey et al. 2007b). 

Among the most important outcomes from these studies was the first ever space-borne search for Supernovae Type Ia in the GOODS fields (Riess et al. 2004). These observations have darker sky background, leading to deeper images, and significantly narrower PSFs, leading to better spatial resolution and hence, identification of more distant supernovae in galaxies, compared to ground-based images. This allowed discovery of 21 high redshift SNe Ia at $z > 1$, which includes almost all (but one) of the highest redshift SNe Ia known at the time (Strolger et al. 2004). Combining these high-$z$~and nearby SNe Ia, the Hubble diagram was established, allowing significant constraints on properties of Dark Energy and its equation of state (Riess et al. 2004, 2007). An essential component of this study was measurement of photometric redshifts of the hosts of SN candidates to identify objects of the highest interest for follow up with subsequent spectroscopic observations. This was possible due to availability of multi-waveband data from space- and ground-based observations. 

The SuperNova Acceleration Probe (SNAP\footnote{http://snap.lbl.gov/}) mission is aimed at finding thousands of SNe of various types to redshift $z\sim$1.7, allowing detailed study of reliability of SNe Ia as standard candles (i.e. dependence of their observed properties on the type of their host galaxy and its redshift, effect of dust, their frequency of appearance in early-type galaxies). This proposed mission will improve the use of of SNe Ia as standard candles and provide significant constraints on properties of Dark Energy and its nature. Moreover, it will provide a large sample of SNe Type II, which are used as diagnostics for star formation activity in galaxies, allowing a statistically large sample of these objects to examine evolution of star formation and metallicity with redshift (Dahlen et al. 2004). As part of the SNAP's primary mission, a wide (at least 1000 square degree per year), multi-band survey will be conducted.  This survey will be optimized for the detection of weak gravitational lensing, a powerful probe of dark energy. Weak lensing provides a direct way for measuring the distribution of dark matter in the Universe. The evolution of dark matter structures over cosmic time is governed by the nature of the dark energy. Thus, accurate photometric redshifts, which are required to measure the 3-dimensional distribution of dark matter, are necessary to exploit weak lensing as a probe of dark energy. Furthermore, these multi-waveband deep data will be extremely useful in studying formation and evolution of galaxies, groups and clusters as a function of their redshift, morphology, environment, color and star formation properties. It will also be a unique mission to probe the first quasars and the first luminous galaxies in the Universe, thus probing the epoch of cosmic reionization.
  
Future Dark Energy probes based on SN Ia and weak lensing ideally require accurate redshifts for individual galaxies. However, given the size of the planned surveys and their depth, it is not practical (nor feasible) to measure spectroscopic redshifts for all the observed galaxies. Therefore, a critical evaluation of the photometric redshift capabilities of any of the JDEM experiments is key to optimize the design of the respective mission. In the case of SNAP, this requires an optimization of the number of filters used, their spectral resolution and shapes and thus throughputs as well as the overall wavelength coverage, to allow most accurate measurement of photometric redshifts. To investigate these problems, we use observational data from the GOODS fields to create a mock galaxy catalog containing a large set of objects with known properties, i.e., redshift, spectral type, luminosity and amount of internal extinction. We then run photometric redshift codes on the multi-waveband data for galaxies in the mock catalog. The aim of this investigation is to study how the accuracy of galaxy photometric redshifts depends on a range of factors including redshift, signal-to-noise, magnitude limits, galaxy spectral type and, in particular, the number of filters, filter shapes and band-widths. For an investigation on how to optimize filters for Type Ia cosmology investigations, see Davis et al. (2006). In a following paper (Jouvel et al. 2007, in preparation), we will investigate in further detail the impact of calibration, SED evolution and size of the spectroscopic surveys on the photomtric redshift determination.

Throughout this paper we use $\Omega_{\Lambda}=0.7$, $\Omega_M=0.3$~and $H_0=70$ km s$^{-1}$ Mpc$^{-1}$. Magnitudes are in the AB system.

\section{The mock galaxy catalog}
\begin{figure*}
\epsscale{0.8}
\plotone{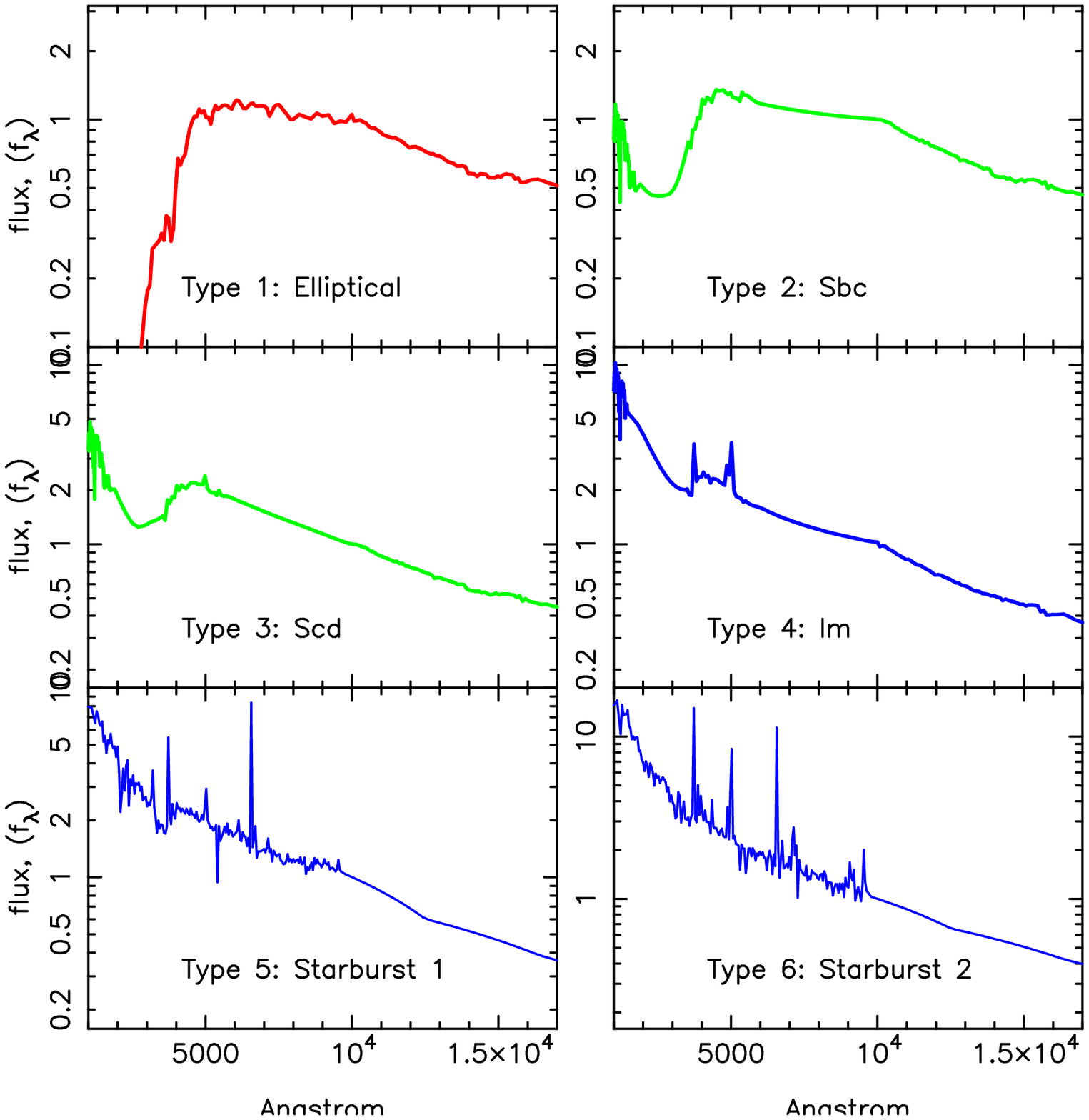}
\caption{The six template SEDs that are used in constructing the mock galaxy catalog. Templates 1-4 are taken from Coleman et al. (1980), while the two starbursts, template 5-6, are taken from Kinney et al. (1996). Templates are extended into UV and near-IR as described in Mobasher et al. (2007).
\label{f1}}
\end{figure*}

To investigate the expected behavior of the photometric redshifts for SNAP, we create a mock galaxy catalog. In the mock catalog, we assign to each galaxy a redshift, a spectral type, an absolute luminosity and a value for extinction. The redshift, spectral type and absolute luminosity are drawn from a distribution according to the observed type-specific luminosity functions (LFs) derived from GOODS (Dahlen et al. 2005). First, a redshift is assigned given by the redshift dependent LF. Second, at the assigned redshift, an absolute magnitude in the range $-24<M_B<-13$~is given to the object according to the LF at that redshift. Finally, the galaxy type is assigned to the object with dependencies on both redshift and absolute magnitude. Redshifts are distributed in the range $0<z<6$. The spectral templates used cover types E, Sbc, Scd, Im (Coleman et al. 1980) and two starburst from Kinney et al (1996, templates SB2 and SB3). The templates are extended in UV and near-IR wave-lengths as described in Mobasher et al. (2007). The template set used is shown in Figure \ref{f1}. To get a continuous set of templates, we make random linear interpolations between adjacent templates when assigning type to the mock galaxies. A random internal extinction is also assigned to each galaxy with a maximum value $E_{B-V}=0.10$~for early types and $E_{B-V}=0.30$~for starbursts. For star forming galaxies, we use a Calzetti et al. (2000) extinction law, while we for later type galaxies assume a Galactic extinction law (Cardelli et al. 1989). We hereafter refer to the redshift in the mock catalog as spectroscopic redshifts.

\begin{figure*}
\epsscale{0.8}
\plotone{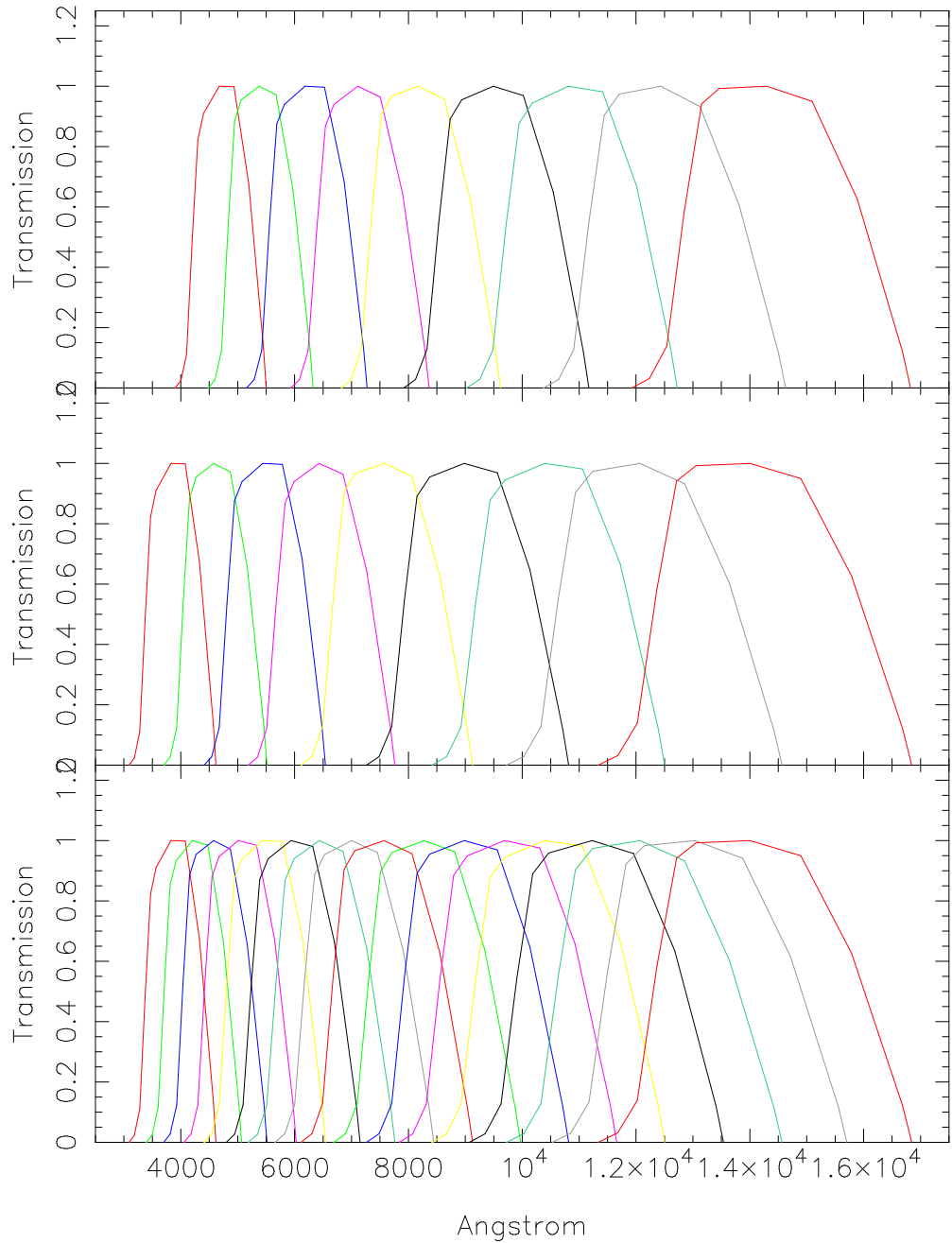}
\caption{Top panel shows the default filter set for SNAP consisting of six optical and three near-IR filters. Filters are indexed 0-8. Mid panel shows the U-band filter set constructed by stretching the standard filter set into the U-band. A 17-filter set (bottom panel) is constructed from the U-band filter set by adding eight intermediate filters. All filters have the peak transmission normalized to unity.
\label{f2}}
\end{figure*}

Using the template SEDs, extinction values, and the response functions for the filter set, we calculate the K-corrections corresponding to the spectroscopic redshift for each galaxy. The K-corrections together with the absolute magnitudes and distance moduli give us the set of apparent magnitudes in each band. To each of these `incident' magnitudes, we add an error to derive the actual `measured' magnitudes that go into the mock galaxy catalog. This `statistical' error is derived from each `incident' magnitude using the S/N at this magnitude (where a S/N=10 corresponds to a magnitude error $\sigma_m\sim$0.10). We thereafter add the error to the magnitude, assuming that errors have a Gaussian probability distribution with $\sigma_m$~as dispersion. This gives the `measured' magnitude. Furthermore, an additional error of 1\% of the flux is also added in quadrature. This accounts for e.g., zero-point uncertainties and photometry uncertainty due to non-perfect image reductions. The default filter set is shown in the top panel of Figure \ref{f2}. This consists of six optical and three near-IR filters, indexed 0-8. Besides the standard set, we also include a `U-band' filter set shown in the mid panel of Figure \ref{f2}. Here we have stretched the standard filter set into the U-band so that the bluest filter has an effective wave-length $\lambda_{eff}$=3910~\AA, compared to the standard filter set which has $\lambda_{eff}$=4750~\AA~for the bluest filter. We include this set to examine the importance of the U-band when it comes to photo-z accuracy and the minimization of the outlier fraction. Furthermore, we include 17-filter set which consists of the same filters as the U-band set with the addition of eight intermediate filters (bottom panel of Figure \ref{f2}). The working concept for the SNAP focal plane is to distribute the six optical detectors and three IR detectors in multiple squares with sizes 6x6 and 3x3, respectively, where each detector has a fixed filter (see Figure 15 in Aldering et al. 2004). This design allows an efficient scanning of all objects in all nine filters. Therefore, it is not possible to add the U-band as a 10:th filter in an efficient way, instead we have to adjust the through-put of the existing filters. For the 17-filter set, the concept is to keep the nominal number of detectors for one of the optical filters, while replacing half of the remaining detectors with the new intermediate filters. This will allow the same scanning advantages, with the exception that 16 of the 17 filters will be half less deep compared to the 9 filter option.

In Table \ref{t1}, we give the limiting magnitudes for S/N=10 and filter characteristics corresponding to the filter sets in Figure \ref{f2}. These are based, with some adjustments, on values given in Aldering et al. (2004), but with fainter limits in the infrared wavelengths, reflecting the recent increase in quantum efficiency for these detectors. To get U-band S/N limit, we assume that the detector quantum efficiency integrated over the U-band is $\sim$65\% of that of the B-band. Note that due to the design of SNAP, the effective exposure time for each of the near-IR bands is twice that of the optical bands, therefore, similar depth (in AB magnitude) is reached over the whole wavelength range. We assume the same total exposure times for all the different scenarios listed in Table \ref{t1}. Therefore, for the cases with more filters, the exposure time per filter is lower, leading the lower S/N ratios. This allows a more meaningful comparison between different scenarios. For the detection filter in the 17-filter set (filter \#11, corresponding to filter \#5 in the 9-filter sets), the exposure time is kept the same as for the other filter sets. This means that a similar number of objects should be detected in all sets when using this band as the detection band. Note that the S/N values should be viewed as representative for a SNAP-like survey, but do not include details about the DQE shapes, mirror reflectivities etc. This will be addressed in Jouvel et al. (in preparation).

\begin{figure*}
\epsscale{0.8}
\plotone{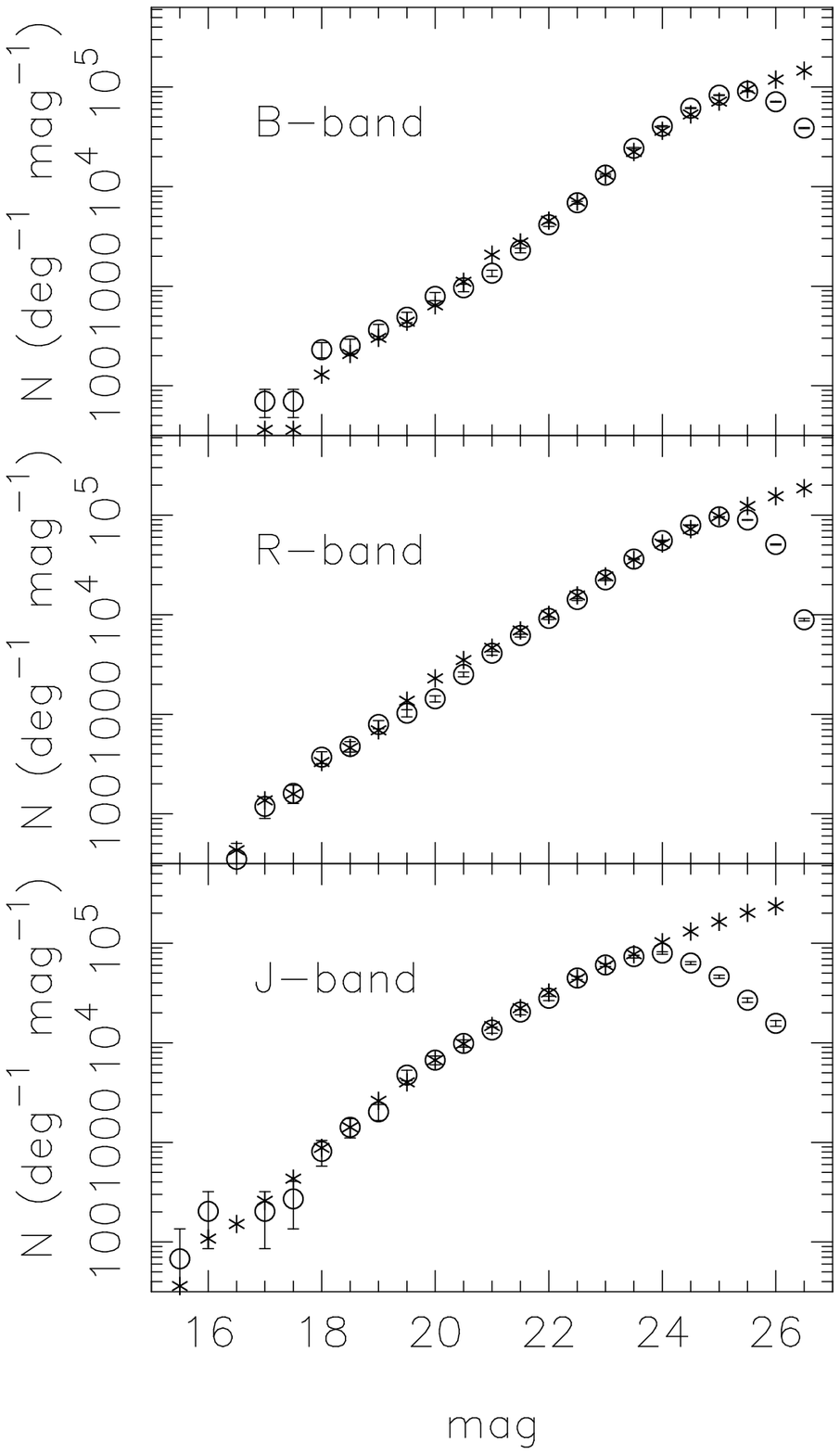}
\caption{Number counts in B-, R-, and J-band for data from GOODS (open circles with statistical error-bars) and a mock galaxy catalog produced using the GOODS filter response functions (asterisk).
\label{f3}}
\end{figure*}

\begin{figure*}
\epsscale{0.8}
\plotone{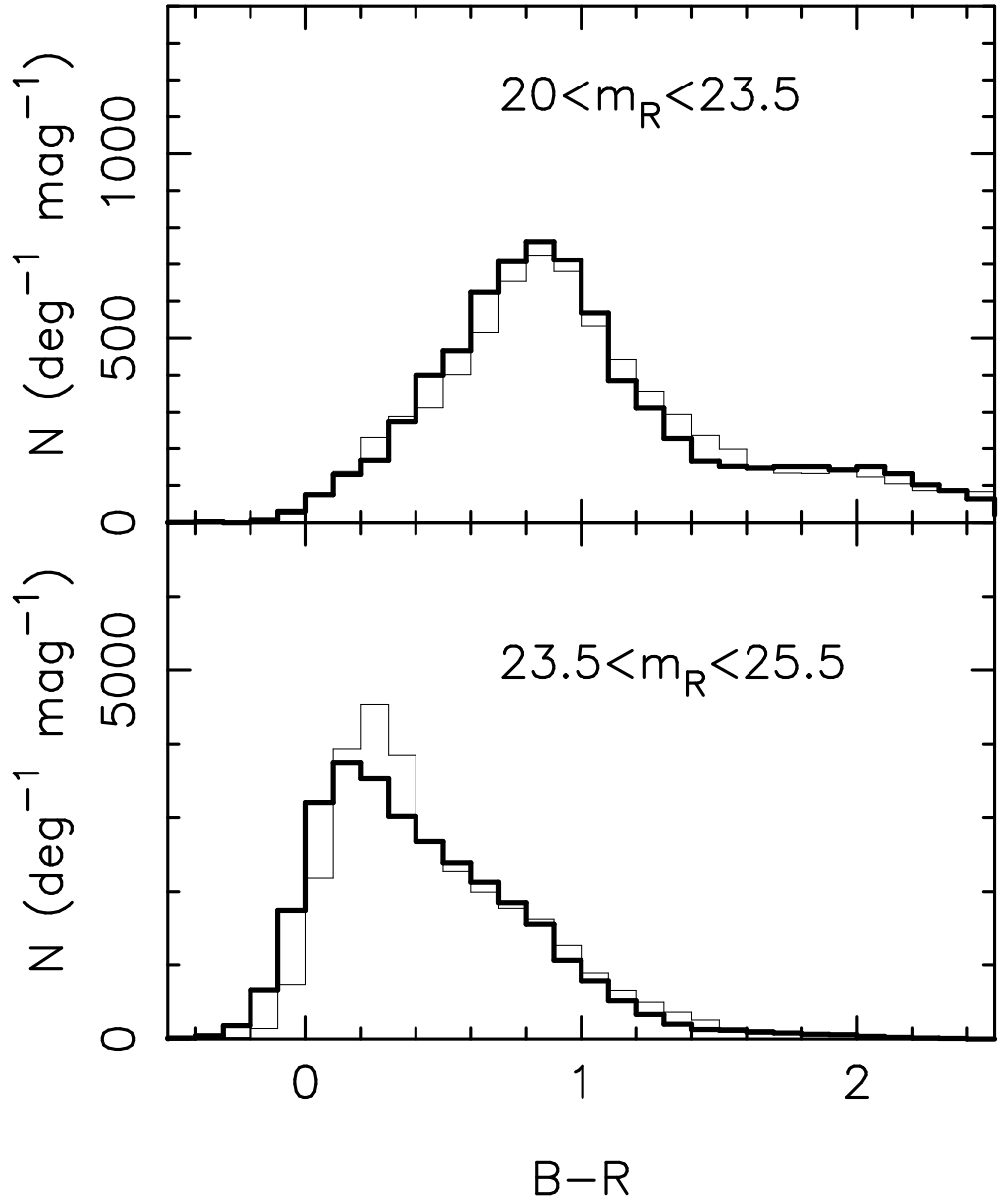}
\caption{Observed B--R for data from GOODS (thick line) and from mock galaxy catalog (thin line). Note the good agreement at both bright and faint magnitudes
\label{f4}}
\end{figure*}

To check how well our code simulates real observed galaxy samples, we produce an additional mock galaxy catalog using the GOODS filter set and compare with the observational data from GOODS-South. Since the mock galaxy catalog is produced using absolute magnitude, spectral type and redshift distributions derived from the GOODS data, a comparison between the observed and simulated galaxy number counts primarily examines if the mock catalog is consistently produced. In Figure \ref{f3}, we show B-, R- and J-band number counts both from the GOODS observations and the mock galaxy catalog. We find that the observed and simulated number counts are in good agreement in all bands. While the mock galaxy catalog is produced from the rest-frame B-band luminosity function, the good agreement in observed B-, R- and J-bands assures us that the mock galaxy colors well represent the colors of the GOODS galaxies. The GOODS data are not as deep as the mock galaxy catalog, making the GOODS counts incomplete at faint magnitudes. To further investigate how the mock galaxy catalog reproduces adequate colors at different magnitudes, we plot in Figure \ref{f4}~the observed and mock galaxy $B-R$~colors in two different magnitude bins. We find a good agreement as well as that the brighter sample has redder colors as expected.

In total, there are $\sim1.6\times10^{6}$ galaxies over an area of 1,000 square arcmin within the specified absolute magnitude and redshift ranges described above. More interesting is the number of galaxies that have an apparent magnitude brighter than a given magnitude limit. In Table \ref{t2}, we give the total number of galaxies as well as the number in ten equally spaced redshift bins to $z<3$. We also give the predicted numbers to $z=6$~in coarser bins. Results are given after applying different S/N cuts for the U-band filter set. The least restrictive cut is S/N$>10$~in-any-filter. Here it suffices that the magnitude in at least one of the filters is brighter than the S/N=10 limits in Table \ref{t1}. The remaining selections are based on the I-band (filter 5), where we choose limits m$_5<$26.6 (corresponding to S/N=10), m$_5<$25.6, and m$_5<$24.6. Note here that the field size chosen results in statistical errors in the number of objects (and in all bins) that are insignificant compared to the uncertainty due to cosmic variance. We estimate that the uncertainty in the GOODS luminosity function due to cosmic variance results in an uncertainty in the numbers in Table \ref{t2} of $\sim$20\%~(Dahlen et al. 2005).

\section{Photometric Redshifts}
To derive photometric redshifts, we use the the template fitting method (e.g., Gwyn 1995; Mobasher at al. 1996). This method compares observed and template SEDs in redshift intervals and assigns photometric redshift and spectral type to individual galaxies by minimizing the $\chi^2$~values

\begin{equation}
\chi^2=\sum^n_{i=1}([F^i_{obs}-\alpha F^i_{template}]/\sigma^i)^2,
\end{equation}
where the summation is taken over the $n$~filters available and $F^i_{obs}$~and $F^i_{template}$~are the observed and template fluxes in band $i$, respectively. Here $F^i_{template}$~includes information on the template SEDs for different galaxy types, absorption values and redshifts, as well as the response curves for the filters. We assume that the latter is well known and do not introduce any additional error exceeding the already included extra errors of 1\%. Finally, $\alpha$~is a normalization constant and $\sigma^i$~is the flux error in band $i$. We use the six template SEDs described above together with two interpolations between each template going from early to later types, making the full set consisting of 16 discrete templates (in contrast to the continuous set in the mock galaxy catalog). Each template SED is redshifted in the range $0<z<6$~in steps $\Delta z=0.01$. We include a luminosity function Bayesian prior in the photo-z fitting. For the prior, we calculate the absolute magnitude the galaxy would have at each tested redshift and compares this with an input luminosity function. If the absolute magnitude corresponding to a particular redshift is improbable, i.e., significantly brighter than $M^*$, then this redshift is disfavored. The input LF used here is independently chosen and is not the same as the LF from which the mock catalogs are generated. The template fitting method and priors we use are further described in Dahlen et al. (2005). We use the same template set for constructing the mock galaxy catalog and calculating photometric redshifts, although the mock galaxies are drawn from a continuous set of galaxies and have had their photometry adjusted by extinction as well as statistical error. Therefore, we can investigate how the photometric redshifts depend on various parameters such as S/N, filter shapes and wavelength coverage, without introducing any bias due to the choice of the templates. However, as a consistency check, we also construct a mock catalog from an alternative set of template SEDs and derive photometric redshifts using the original templates in the fitting. This is described in \S 5.4.

\section{Results}
We present results on the accuracy of the photometric redshifts after applying different magnitude cuts as discussed above. The accuracy of the photometric redshifts, $\sigma_z$, is defined as:
\begin{equation}
\sigma_z\equiv rms[(z_{spec}-z_{phot})/(1 + z_{spec})].
\end{equation}
For the results presented here we focus on a redshift range $0<z_{phot}<3$, while the redshift range of the mock galaxy catalog is $0<z_{spec}<6$. Therefore, any galaxy with true redshift $z>3$~that is scattered into the $z_{phot}<3$~range, will be included. This is important for deriving the correct redshift errors for the photometric redshift selected sample. Note that at $z>6$, the galaxy light is redshifted longward of 8000\AA~and galaxies will only be detected in the four reddest filters and will therefore not be part of the primary lensing catalog. Furthermore, the shapes of these galaxies will not be well measured because of the coarser pixels of the NIR detectors and because they are expected to be extremely faint. We present results for the full sample of galaxies in the mock catalog to specified magnitude limits, together with results after rejecting outliers with large errors, so called `catastrophic failures'. The overall accuracy of the photometric redshifts is often dramatically increased after excluding outliers. We therefore also quote the results after excluding outliers, together with the fraction of these objects. Outliers are here defined as objects with $|(z_{spec}-z_{phot})/(1 + z_{spec})|>0.3$. Note, however, that in a real situation it is not possible to know which galaxies are outliers. Later we discuss methods for identifying galaxies with a reasonable probability of being outliers.

In Table \ref{t3}, we present results, using four different magnitude cuts and the three filter sets discussed. The quoted values for $\sigma_z$~and outlier fraction are subject to a statistical uncertainty depending on the mock catalog sample size. Using simulations we find that the uncertainties in $\sigma_z$~and the outlier fractions are $<0.2$\%~and $<0.04$\%, respectively. The uncertainty in the number densities is $\gsim$20\%, mainly caused by cosmic variance.

\begin{figure*}
\epsscale{0.8}
\plotone{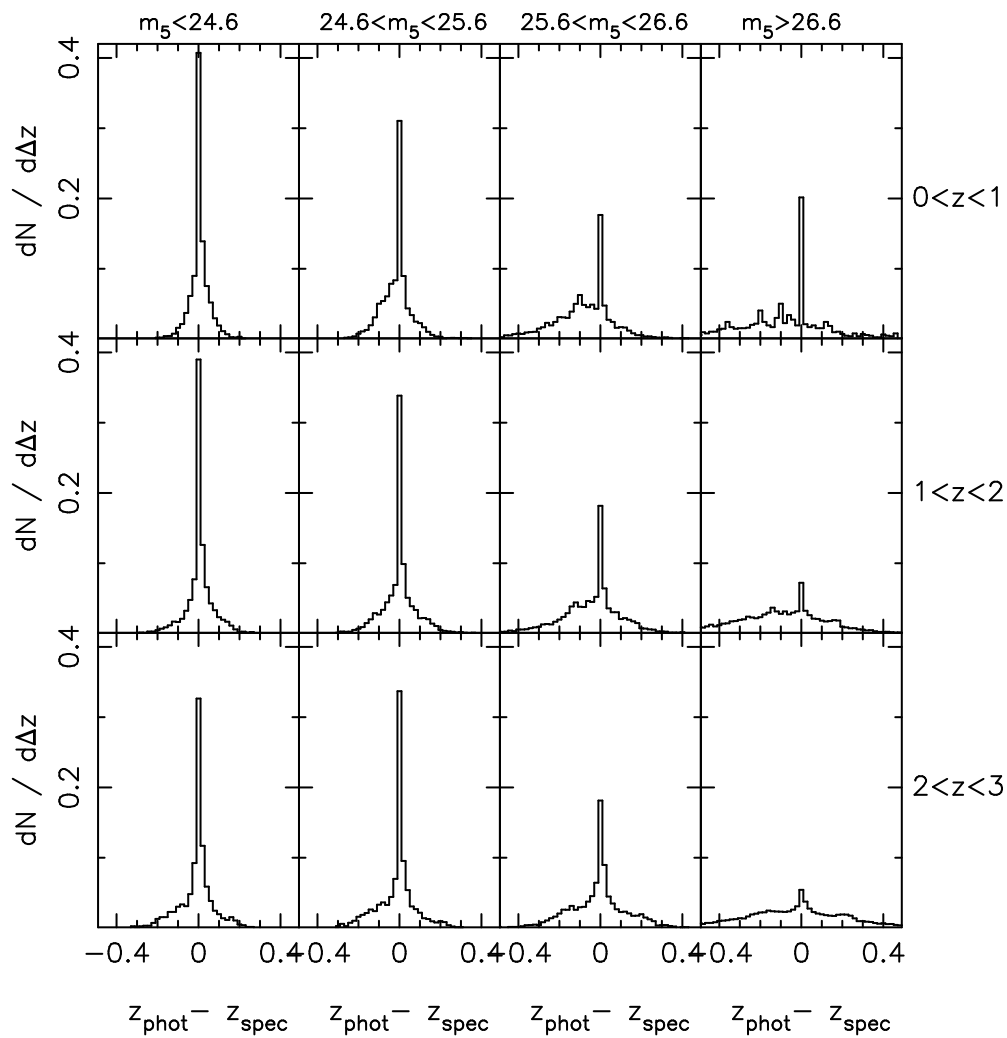}
\caption{Distributions of photometric minus simulated ``spectroscopic'' redshifts for different magnitude bins. Going from left to right, the cuts are m$_5<$24.6, 24.6$<$ m$_5<$25.6, 25.6$<$ m$_5<$26.6 and m$_5>$26.6 combined with the requirement that S/N$>10$~in at least one of the remaining bands. Distributions are normalized to unity.  
\label{f5}}
\end{figure*}

As expected, the accuracy in the photometric redshifts increases when using brighter magnitude cuts. At the same time, this naturally also reduces the number of objects for which photometric redshifts are calculated (Table \ref{t3}). The increased scatter at fainter magnitude cuts is also evident in Figure \ref{f5}, where we plot distributions of the photometric minus spectroscopic redshifts for the U-band filter set in different magnitude bins. Going from the brightest magnitude bins (left panels) to successively fainter cuts (right panels) we note an increased scatter. There is also a slight increase in scatter at higher redshifts (comparing top panels with bottom panels), but not as large as the dependence on magnitude. Furthermore, the figure clearly illustrates the non-Gaussian shape of the error distribution of the photometric redshifts.

Comparing the results for the different filter sets in Table \ref{t3}~reveals that the accuracy increases when including the U-band. However, the most important difference is the clear reduction in outlier fraction when including the U-band. For example, at m$_5<$26.6, the outlier fraction decreases by a third when including the U-band, while at m$_5<$25.6 the difference is more than a factor four.  This can be attributed to the ability of the U-band to better probe the 4000\AA-break at low redshifts and the Lyman-break at redshifts close to $z\sim 3$. The B-band probes the Lyman-break at $z\gsim 3$, and is therefore not helpful in the redshift range investigated here. 

Since the numbers of detected objects are similar for the standard and the U-band filter sets, there are only advantages in extending the wave-length coverage to the U-band, assuming that the S/N values, in particular for filter \#0, are representative. Therefore, we hereafter concentrate on the U-band filter set. The resulting rms for the U-band filter set is in the range $\sigma_z \sim 0.03-0.14$, depending on selection. However, most of the spread, especially at the fainter cut, is mainly due to a few objects enhancing the errors. As can be seen from the table, after excluding a few per cent or less of objects classified as outliers, the rms drops to $\sigma_z \sim 0.03-0.07$. Below, we discuss methods for identifying objects that are outlier candidates. 

It should here be emphasized that the inclusion of the near-IR filters is essential for the accuracy of the photometric redshifts and the minimization of the outlier fraction, especially when aiming at redshifts $z>1$, where the rest-frame 4000\AA-break moves out of the optical bands. Running our photometric redshift code using the U-band filter set, but excluding the three near IR filters, we find an increase in the fraction of outliers by a factor $\sim$10, somewhat depending on selection. It is thus very important to have both U and near-IR filters to obtain high quality photometric redshifts.

\begin{figure*}
\epsscale{0.7}
\plotone{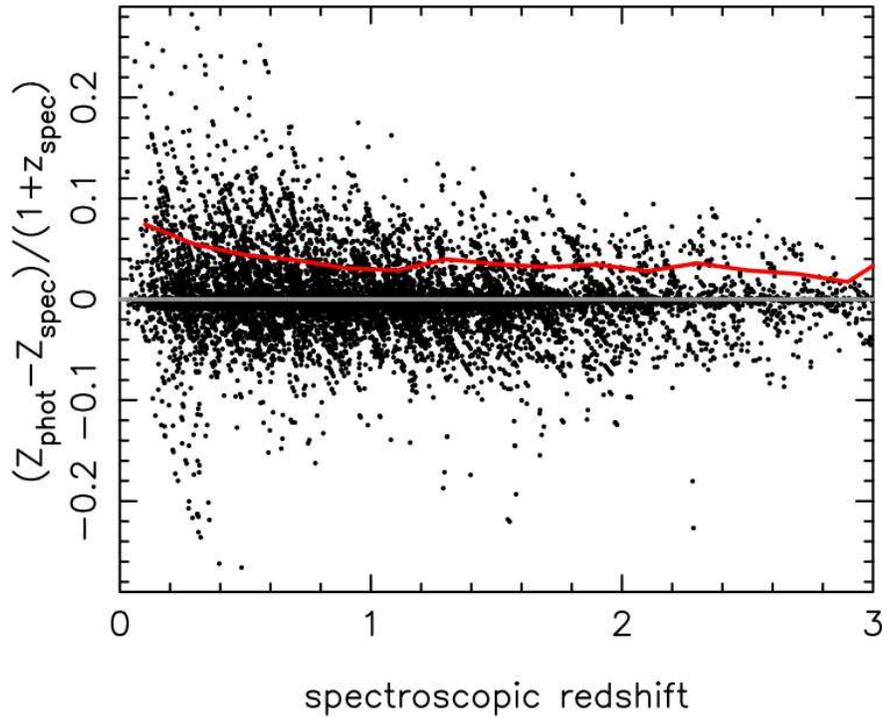}
\caption{Scatter between photometric redshift and spectroscopic redshift as a function of spectroscopic redshift, using U-band filter set and m$_5<25.6$. Red line shows changes in the rms. A flat change in rms with redshift indicate that photometric redshift errors scale proportional to (1+$z$).
\label{f6}}
\end{figure*}

\begin{figure*}
\epsscale{0.7}
\plotone{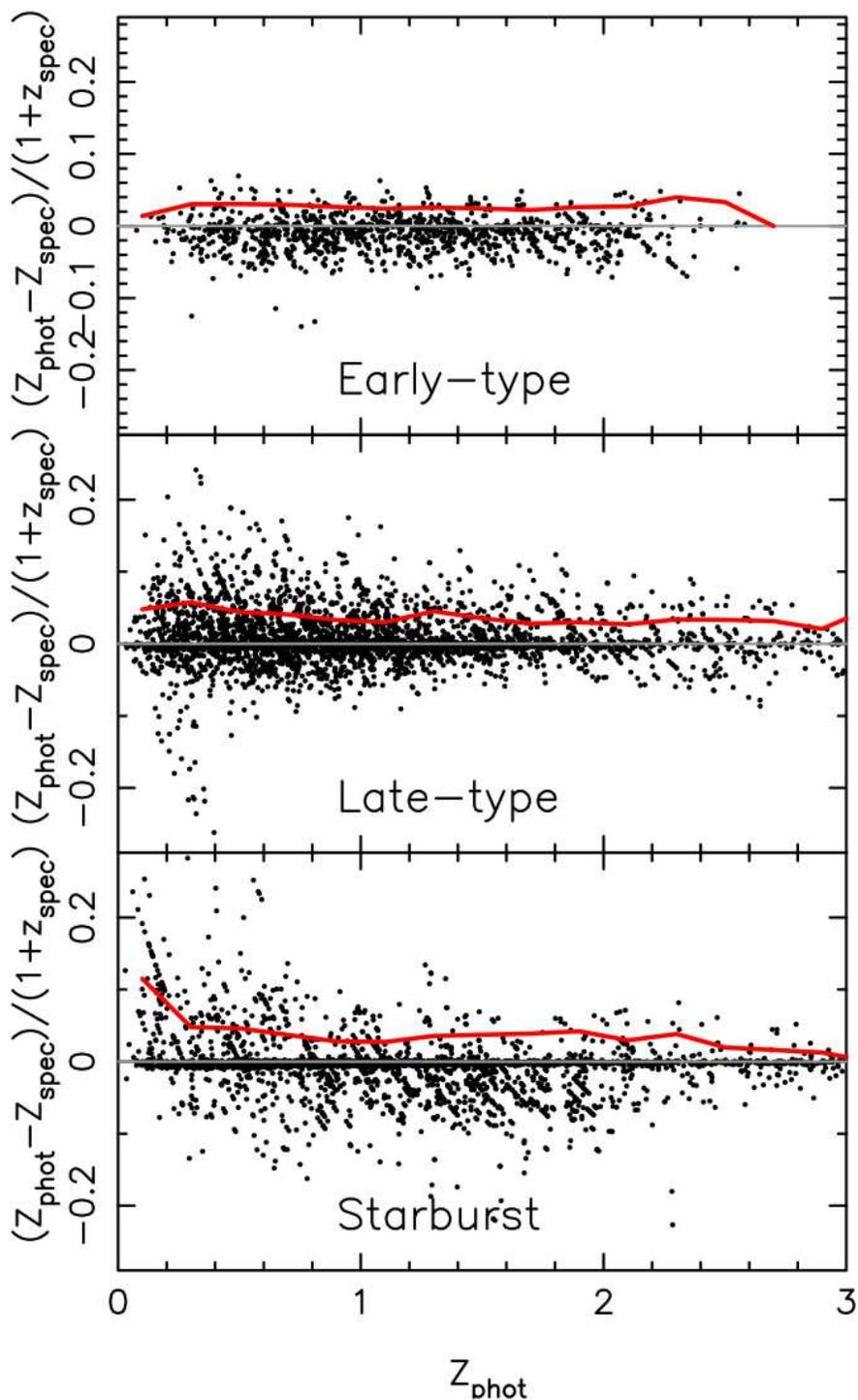}
\caption{Scatter between photometric redshift and spectroscopic redshift as a function of spectroscopic redshift, using U-band filter set and m$_5<25.6$. Top panel shows results for early-type galaxies, while mid and bottom panels show late-type galaxies and starbursts, respectively. Red lines show changes in the rms.
\label{f7}}
\end{figure*}

Next we investigate in some more detail how the accuracy of the photometric redshifts depends on redshift. Figure \ref{f6} shows the normalized difference between photometric and spectroscopic redshifts ([$z_{phot}-z_{spec}$]/[1+$z_{spec}$]) to redshift $z=3$. The red line shows changes in {\it rms} with redshift. The small variation of the rms with redshift shows that expected redshift errors scale as $\sim$(1+$z$), confirming that using normalized errors gives a fair ``redshift independent'' measurement of the scatter. In Figure \ref{f7}, we divide the sample into different spectral types as given by the best-fitting template SEDs. The figures clearly show that the early-type galaxies have a smaller fraction of galaxies with high scatter. This is also shown in Table \ref{t3}~where earlier type galaxies have lower scatter and outlier fractions. We expect this behavior since earlier types have the strongest 4000\AA-break, the most important spectral feature for determining photometric redshifts at $z<3$.

\begin{figure*}
\epsscale{0.7}
\plotone{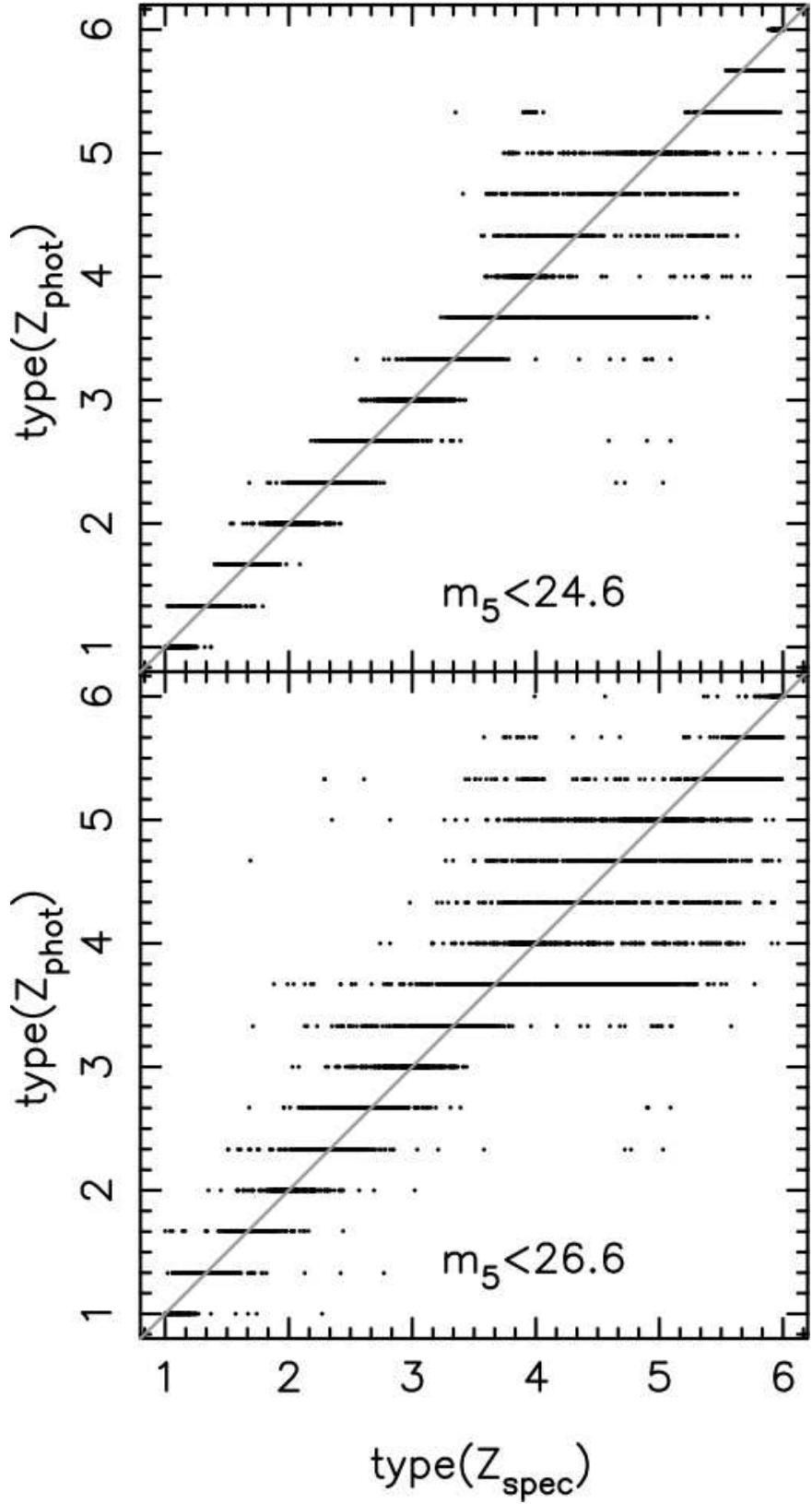}
\caption{Comparison between input ``spectroscopic'' spectral type and derived
photometric spectral type for two different magnitude limits. 
\label{f8}}
\end{figure*}

The photometric redshift code returns the best-fitting spectral template for each galaxy, and since we know the input spectral type, we can estimate the accuracy in deriving galaxy spectral types. Figure \ref{f8}~shows the scatter between input and derived spectral types at different magnitude cuts, with the spectral types numbered from 1 to 6, according to Figure \ref{f1}. As expected, there is a better agreement between input and derived spectral type for brighter magnitudes. Also, the scatter is larger for later type galaxies, mainly due to the relative similarity between the colors of types 4-6. For the m$_5<26.6$~and m$_5<24.6$~cuts, we find that respectively 92\% and 94\%~of the galaxies are given a spectral type within $\pm$1 from the input type. Note that the quantized structure of Figure \ref{f8} is caused by the discrete set of 16 template SEDs used in the photometric redshift method.

\subsection{Redshift $3<z<6$~interval}
While the SNAP mission is foremost focussed on the redshift range $z<3$, a large number of higher redshift objects will be detected. In Table \ref{t2}, we give the predicted number counts to $z=6$, however, since these counts are derived from extrapolations of lower redshift LFs, the uncertainty is high. Investigating the photometric redshifts using the U-band filter set, we find a significant increase in the outlier fraction at $z_{phot}>3$, in particular at faint magnitudes. At m$_5<26.6$, we estimate $\sim$30\% outliers, decreasing to $\sim$4\%~and $\sim$1\%~at  m$_5<25.6$~and m$_5<24.6$~in the redshift range $3<z<6$, respectively. Excluding the outliers, the accuracy is comparable to the lower redshift case. These results are presented in Table \ref{t3}.

\section{Discussion}
\subsection{Reducing fraction of outliers: D95 method}
As already noted, a few outliers with ``catastrophic redshifts'' are often responsible for a large part of the estimated scatter between spectroscopic and photometric redshifts. Therefore, it is valuable to have methods for finding and flagging objects that may be outliers, which thereafter could be excluded. A successful method should identify as many outliers as possible, while keeping the total number excluded at a minimum. In Mobasher et al. (2006), we discussed and proved the utility of the so called D95-method. We define D95 as the width of the 95\% confidence interval derived from the photometric redshift fitting divided by one plus the photometric redshift,
\begin{equation}
{\rm D95=95\%~confidence~interval/(1+}z_{phot}).
\end{equation}
Large D95 values are assigned to galaxies with wide 95\% confidence intervals. These could be due to a broad peak in redshift probability distribution caused by large photometric errors, or it could reflect a double peak in the probability distribution. The latter case happens when there is a confusion between the Lyman-break and the 4000\AA-break. Therefore, when D95 is large, the uncertainty in the photometric redshift will also be large, increasing the probability that the galaxy is an outlier.

In Table \ref{t3}, we give results on the photometric redshift accuracy after applying different cuts in D95 for the U-filter set. The results show that it is possible to significantly reduce the fraction of outliers, while only decreasing the total fraction of objects by a small amount. For example, for the ``S/N $>$ 10 in-any-filter'' selection and a D95$<$0.40 cut, the fraction of outliers decreases by 95\%, while the total number of objects decreases by 34\%. For the m$_5<$26.6 selection, a cut D95$<$0.40 decreases the number of outliers by 90\%, while only decreasing the total number of objects by 14\%.

We have previously shown that applying a brighter magnitude cut also decreases the number of outliers and, at the same time, decreases the number of galaxies in the sample. Inspecting Table \ref{t3} shows that better results are obtained using the D95 method when requiring a particular number density of galaxies. For example, using the ``S/N $>$ 10 in-any-filter'' selection together with a D95$<$0.40 cut compared to using the m$_5<$26.6 selection without a D95 cut, results in a similar number of objects but with better photometric redshifts and significantly fewer outliers for the former selection. It therefore seems more efficient to use D95 as the primary criterion when making a cut in the galaxy sample to improve photometric redshift accuracy compared to using a magnitude cut.
 
\subsection{Bias}
Weak-lensing studies show that a small photometric redshift bias is important for accurate results. The bias is here defined as the mean offset between the photometric redshifts and the true ``spectroscopic'' redshift. Typically, a bias $<$0.003(1+$z$) in each of ten redshift bins to $z$=3 is desirable, assuming the survey is reaches a depth where at least 100 galaxies per square arcmin have determined photometric redshifts. (Ma et al., 2006; Huterer et al. 2006). 

\begin{figure*}
\epsscale{0.7}
\plotone{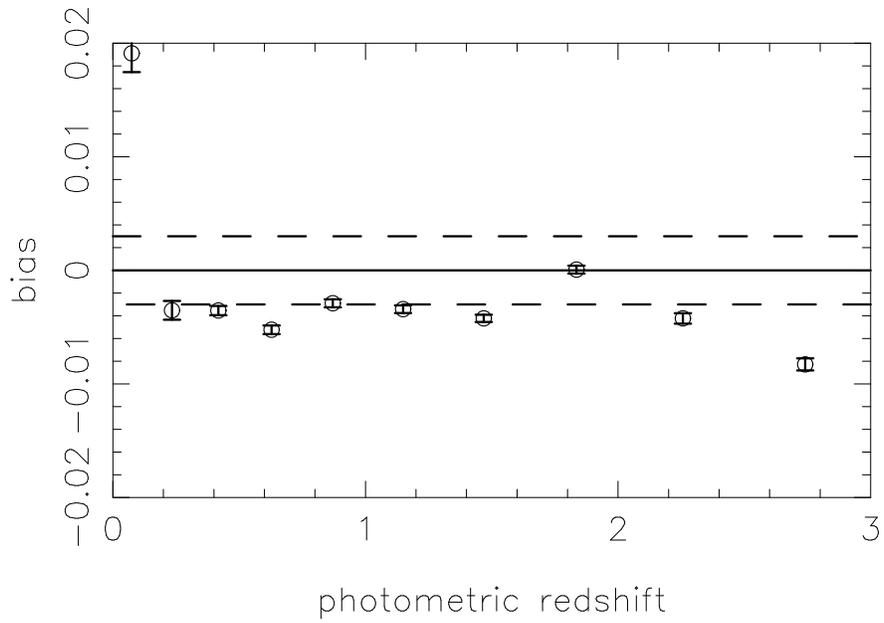}
\caption{Normalized bias (mean offset between photo-z and spec-z divided by (1+$z$)) in 
10 redshift bins with widths ln(1+$z$)=0.14. The sample shown is selected by m$_5<$26.6 
and D95$<$0.3 and has a galaxy density $\sim$90 per sq arcmin to $z<3$. Dashed lines
shows the bias$<$0.003(1+$z$) desired limit.
\label{f9}}
\end{figure*}

In Figure \ref{f9}, we plot the bias in ten redshift bins to $z$=3 for a magnitude limit m$_5<$26.6 and D95$<$0.3, including $\sim$90 galaxies per sq arcmin. Bin size is chosen so that ln(1+$z$)=constant. It is clear from the figure that the results are near the desired value (dotted lines) in most bins. This is reassuring since the aim of this investigation is not to minimize bias. In a real situation, using a training set of galaxies with spectroscopic redshifts should allow to minimize the bias.  What the Figure tells us is that the bias offsets may be a problem at the very lowest and highest redshifts, which could be due to low statistics and relatively high outlier fractions.

\subsection{Alternative filter sets}
\subsubsection{A 17-filter set}
In our simulations we also include a 17-filter set. The rationale behind this is that with a narrower spacing in wave-length between filters, it should be easier to pick up the location of the redshifted 4000\AA-break, and therefore determine the redshift with higher accuracy. The drawback is that with a fixed amount of observing time available, the S/N in each individual filter decreases. In our simulations, we have kept the nominal exposure time in the detection filter, while decreasing the exposure time in the remaining 16 filters by a factor two (consistent with the detector configuration discussed in \S 2).

Results using the 17-filter set is presented in Table \ref{t3}. Compared to the 9-filter sets, the 17-filter set has a larger scatter except at the brightest magnitudes. However, after excluding outliers, the 17-filter set shows smaller scatter at all magnitudes. Both sets have comparable number of outliers. We therefore conclude that the 9-filter set is preferred except at the brightest magnitudes and that with a method that efficiently excludes outliers, the 17-filter set should be comparable, or even better, than the 9-filter set.

\subsubsection{Less wide broad-bands}
The filters used so far are fairly broad with significant overlap between them (see Figure \ref{f2}). As an alternative, we also construct a set that has the same effective wave-lengths as the U-band filter set, but with filter widths being only 75\% of the original. This is approximately equivalent of changing the resolution from $\sim 4$~to $\sim 5.3$~(numbers are somewhat filter dependent). The reason for testing this alternative filter set is to investigate if the increased resolution will make it easier to locate the redshifted spectral breaks in the galaxies' SEDs and therefore decrease the scatter in the photometric redshifts. Our results show that there is no gain in the accuracy due to the resolution, but instead, the scatter in the photometric redshifts increases due to the larger errors caused by lower counts in the narrower filters. At the faintest limits (m$_5<26.6$), both the scatter and fraction of outliers is twice that of the full width broad band filter set. At bright magnitudes (m$_5<$25), both sets give comparable results. Also, with the narrower filters, the number of objects with S/N$>$10 decreases by $\sim$ 14 \%.

\subsubsection{Square filters}
We finally include a filter set with square transmission functions. These filters have, by construction, the same area (i.e., integral of transmission over wave-length) as the U-band filter set and are centered on the effective wave-lengths of those filters. This leads to a filter set with similar resolution compared to the U-band set. The resulting photometric redshifts show three differences compared to the U-band filter set. First, the scatter (including outliers) increases for all magnitude selections. Second, the fraction of outliers increases by a factor $\sim$2 at the faintest selections, while being similar at brighter magnitudes (m$_5<25.6$). Finally, the scatter in the photometric redshifts decreases after excluding the outliers (down $\sim$15-30\%) at all selections. Overall, this mean that the photometric redshifts for most objects is somewhat improved, but for a few objects there is a large degradation in the accuracy. Therefore, for the best overall accuracy, the standard filter set gives the best results, however, the difference is quite marginal.
\subsection{Investigating an alternative set of SEDs}
So far in this investigation, we have used the same set of template SEDs when creating both the mock galaxy catalog as when deriving photometric redshifts. This allows us to concentrate on how the photometric redshift accuracy depends on S/N and filter choices without adding biases that can be introduced if different galaxy sets are used for creating catalogs and deriving redshifts. However, the real case will be different from this investigation in the sense that it will not be known a priori if the template set used for calculating the photometric redshifts represents the actual distribution of SEDs for the real galaxies. 

\begin{figure*}
\epsscale{1.0}
\plotone{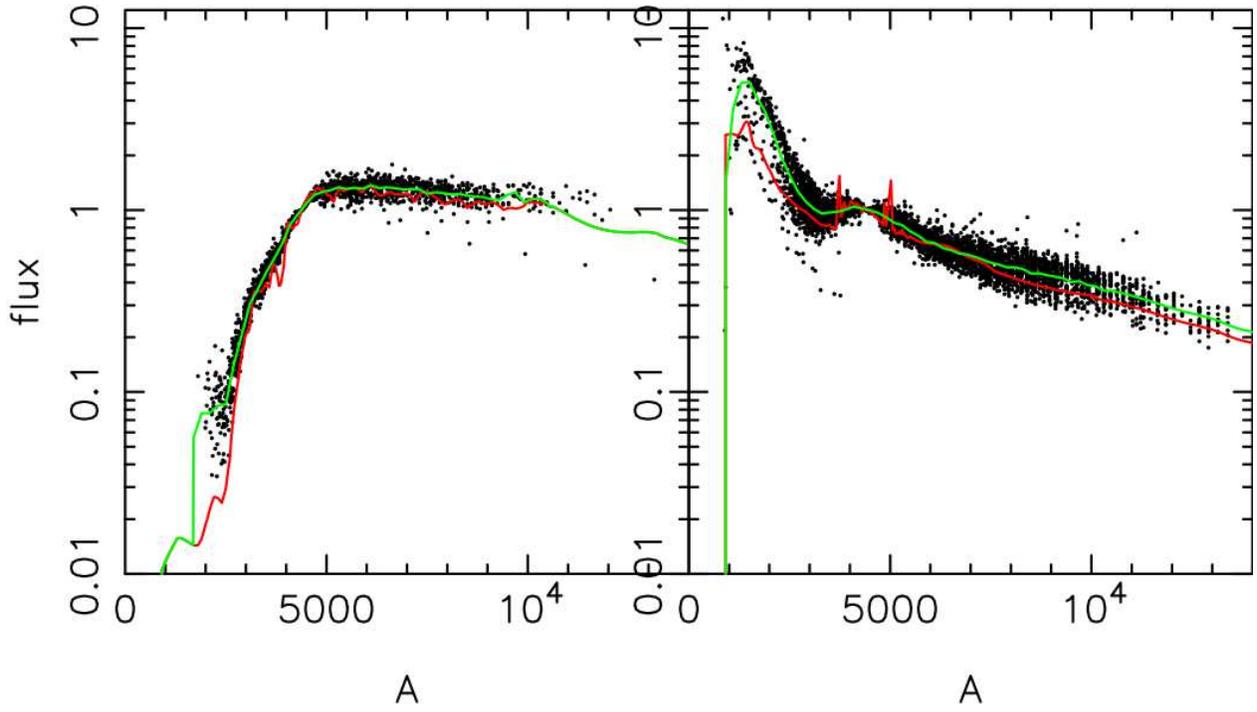}
\caption{Correcting template SEDs using spectroscopic sample. For each
galaxy, the observed flux is normalized to rest-frame 4400\AA. Red lines show the
input template SED, while green lines show the corrected SED after fitting
to the data. Left panel shows an early-type galaxy, while right panel shows
an Im type galaxy.
\label{f10}}
\end{figure*}

To investigate how well we can derive photometric redshifts in a situation where do not know the shape of the observed galaxies SEDs, we made a new set of simulations with the intent of calibrating our template SEDs using spectroscopic redshifts, an approach shown successful by Ilbert et al (2006). We create an alternative mock galaxy catalog using a different set of galaxy template SEDs. This second set of template SED are based on the PEGASE galaxy models (Fioc \& Rocca-Volmerange 1997) and includes six templates from Elliptical to Starburst. We also added systematic offsets to the galaxy magnitudes (of order a few 0.01 mag) to mimic zero-point calibration uncertainties, and dust extinction. For a subset 'training sample' of 10$^4$~galaxies, the true spectroscopic redshift was given to test and calibrate the photometric redshifts. Photometric redshifts were thereafter derived using the first set of template SEDs and compared to the spectroscopic sample. The first run produced a fairly large scatter in the redshifts due to both the zero-point offsets and SED mismatches. To decrease scatter we use two approaches. First we add offsets to the catalog magnitudes and rerun the photometric redshift code minimizing the scatter between photometric redshifts and the spectroscopic redshifts in the training sample. Second, a new set of modified template SEDs was created. To make this, we first divided the spectroscopic sample into six types using the template set from the photometric redshift code (without corrections for dust extinction). For each type, we thereafter plotted the flux of all objects at the rest frame wave-length of each filter normalized to 4400~\AA. Figure \ref{f10} shows the case for early-type galaxies (left) and Im type galaxies (right). Red lines show the original template SED, while green lines show a fit to the data. We adopt these fits as a new set of corrected template SEDs. We finally recalculate the photometric redshifts using both zero-point corrections and corrected template SEDs. The results are consistent with the results given in Table \ref{t3} both in terms of scatter and fraction of outliers, assuring us the size of the scatter presented here will not dramatically change even though the SEDs of the actual galaxies observed are not the same as the assumed set of template SEDs. Note, however, that the accuracy of the photometric redshifts will depend on the spectroscopic sample and the diversity of the SED population. The more diverse the SEDs of the true galaxy population are, the larger the number of spectroscopic redshifts needed. This will be further investigated Jouvel et al. (in preparation).

\section{Conclusions and Summary}
We have used simulations to investigate how the behavior of photometric redshifts for a SNAP-like mission depends on e.g., magnitude limits, filters choices, wavelength coverage and galaxy types. We have also discussed methods for decreasing the expected fraction of outliers, i.e., galaxies with significant disagreement between the photometric and spectroscopic redshifts. We note that our investigation is not primarily focussed on getting exact numbers for e.g., the photometric redshift accuracy and outlier fractions, but to investigate how these diagnostics are affected when changing e.g., S/N and filter sets. Our main conclusions are: 
\begin{itemize}
\item{We find that including the U-band significantly decreases the fraction of outliers and results in an increase photometric redshift accuracy.}
\item{A 17-filter set results in larger scatter compared to a 9-filter set except at bright magnitudes. However, after excluding outliers, the 17-filter set gives more accurate redshifts at all magnitudes.}
\item{A 9-filter set with narrower filter functions (resolution $\sim$5 instead of $\sim$4), results in an increase in the scatter of the photometric redshifts. This is caused by the larger photometric errors when fewer photons are detected.}
\item{Using a filters with a square transmission curves  decreases the scatter in the photometric redshifts for the majority of the objects. At the same time, however, the fraction of outliers is doubled. Therefore, if possible outliers could be efficiently flagged, the square filter set would be preferred.}  
\item{The accuracy of the photometric redshifts depends on both magnitude (or efficiently the S/N) and galaxy spectral type, with better results at high S/N and for earlier type galaxies.}
\item{Using the D95 method can significantly decrease the number of outliers, while only decreasing the total number of objects moderately. Using this method is therefore preferred compared to using only the S/N as a cut to decrease scatter.}
\end{itemize}

\acknowledgments{
}

\clearpage
\begin{deluxetable}{ccrccccccc}
\tabletypesize{\scriptsize}
\tablewidth{0pt}
\tablecaption{S/N=10 limiting magnitudes}
\tablecolumns{10}
\tablehead{
\colhead{Filter}& \colhead{Standard} & \colhead{$\lambda_{eff}$} & \colhead{$\Delta\lambda$}& \colhead{U-band set} & \colhead{$\lambda_{eff}$} & \colhead{$\Delta\lambda$}& \colhead{17-band set} & \colhead{$\lambda_{eff}$} & \colhead{$\Delta\lambda$}\\
}
\startdata
0  & 26.8 &  4750 & 1070 & 26.6 & 3950 & 990 & 26.2 & 3950 & 990 \\
1  & 26.7 &  5450 & 1240 & 26.7 & 4730 & 1180 & 26.3 & 4340 & 1090 \\
2  & 26.6 &  6270 & 1440 & 26.6 & 5610 & 1410 & 26.3 & 4730 & 1180 \\
3  & 26.6 &  7200 & 1630 & 26.6 & 6640 & 1660 & 26.2 & 5170 & 1300 \\
4  & 26.6 &  8270 & 1870 & 26.6 & 7820 & 1950 & 26.2 & 5610 & 1410 \\
5  & 26.6 &  9610 & 2190 & 26.6 & 9270 & 2320 & 26.2 & 6130 & 1540 \\
6  & 26.7 & 10960 & 2500 & 26.7 & 10730 & 2680 & 26.2 & 6640 & 1660\\
7  & 26.7 & 12570 & 2820 & 26.7 & 12460 & 3110 & 26.2 & 7230 & 1810\\
8  & 26.7 & 14450 & 3300 & 26.7 & 14450 & 3620 & 26.2 & 7820 & 1950\\
9  &      &  & &     & & & 26.2 & 8540 & 2140\\
10 &      &  & &    & & & 26.6 & 9270 & 2320\\
11 &      &  & &    & & & 26.3 & 10000 & 2500\\
12 &      &  & &    & & & 26.3 & 10730 & 2680\\
13 &      &  & &    & & & 26.3 & 11590 & 2900\\
14 &      &  & &    & & & 26.3 & 12460 & 3110\\
15 &      &  & &    & & & 26.3 & 13450 & 3360\\
16 &      &  & &    & & & 26.3 & 14450 & 3620
\enddata
\tablecomments{Indicative S/N=10 limits for SNAP filters based on values from Aldering et al. (2004). The U-band filter set is constructed by stretching the standard filter set to shorter wave-lengths, while the 17-filter set is based on the U-band set with eight additional intermediate filters. Due to the stretching, the latter two sets have somewhat wider filters (lower resolution) compared to the standard set. Filter sets used are shown in Figure \ref{f2}.
}
\label{t1}
\end{deluxetable}
\clearpage
\begin{deluxetable}{lcccccc}
\tabletypesize{\scriptsize}
\tablewidth{0pt}
\tablecaption{Galaxy Number Counts}
\tablecolumns{5}
\tablehead{
\colhead{Redshift} & \colhead{S/N$>$10 in-any}$^1$ & \colhead{m$_5<$26.6} & \colhead{m$_5<$25.6} & \colhead{m$_5<$24.6}
}
\startdata
\multicolumn{5}{c}{Number of galaxies per sq. arcmin}\\
\hline
$0.0<z<3.0$   & 156 &  110 & 67 & 36  \\
$0.0<z<0.3$   & 5.0 & 5.0 &  4.8 &  3.7  \\
$0.3<z<0.6$   &  15 &  15 & 12 &  8.0  \\
$0.6<z<0.9$   &  21 &  20 & 13 &  8.0  \\
$0.9<z<1.2$   &  23 &  21 & 12 &  6.9  \\
$1.2<z<1.5$   &  23 &  17 &  9.2&  4.8  \\
$1.5<z<1.8$   &  24 &  13 &  6.5 &  2.8  \\
$1.8<z<2.1$   &  15 &   7.3 &  3.6 &  1.3  \\
$2.1<z<2.4$   &  14 &   5.4 &  2.2 &  0.54  \\
$2.4<z<2.7$   &   9.8 &   4.1 &  1.4 &  0.27  \\
$2.7<z<3.0$   &   7.0 &   3.1 &  0.96 &  0.14\\
\hline
$3.0<z<4.0$   &  16 &   7.3 &  1.4 &  0.14  \\
$4.0<z<5.0$   &   12 &   4.6 &  0.41 &  0.01  \\
$5.0<z<6.0$   &   4.2 &   1.3 &  0.10 &  0.00

\enddata
\tablecomments{1.) The selection requires that the magnitude in 
at least one filter is brighter 
then the S/N values given in Table \ref{t1}~for the U-band filter set. Other filter
sets should have comparable numbers, assuming that S/N in detection band is the same.
Clustering variance may add uncertainty of at least $\sim$20\%. Numbers at $z>3$~are more 
uncertain since they are derived using extrapolations of lower redshift luminosity functions. 
}
\label{t2}
\end{deluxetable}

\clearpage
\begin{deluxetable}{lcccccc}
\tabletypesize{\scriptsize}
\tablewidth{0pt}
\tablecaption{Results}
\tablecolumns{5}
\tablehead{
\colhead{Sample} & \colhead{$\sigma$ z} & \colhead{$\sigma$ z w/o outliers} & \colhead{fraction outliers} & N/sq arcmin 
}
\startdata
\multicolumn{5}{c}{Standard filter set, $0<z<3$}\\
S/N $>$ 10 in-any filter & 0.135 & 0.073 & 0.0361 & 156\\
m$_5<$ 26.6              & 0.096 & 0.060 & 0.0183 & 113\\
m$_5<$ 25.6              & 0.053 & 0.044 & 0.0032 & 70\\
m$_5<$ 24.6              & 0.041 & 0.038 & 0.0017 & 39\\
m$_5<$ 25.6, Early       & 0.040 & 0.039 & 0.0003 & 7\\
m$_5<$ 25.6, Late        & 0.042 & 0.039 & 0.0009 & 38\\
m$_5<$ 25.6, Starburst   & 0.068 & 0.051 & 0.0075 & 25\\
\hline
\multicolumn{5}{c}{U-band filter set, $0<z<3$}\\
S/N $>$ 10 in-any filter & 0.140 & 0.071 & 0.0320 & 156\\
m$_5<$ 26.6              & 0.090 & 0.055 & 0.0120 & 110\\
m$_5<$ 25.6              & 0.041 & 0.037 & 0.0007 & 67\\
m$_5<$ 24.6              & 0.033 & 0.032 & 0.0003 & 36\\
m$_5<$ 25.6, Early       & 0.031 & 0.028 & 0.0003 & 6\\
m$_5<$ 25.6, Late        & 0.041 & 0.038 & 0.0005 & 37\\
m$_5<$ 25.6, Starburst   & 0.042 & 0.038 & 0.0011 & 24\\
\multicolumn{5}{c}{U-band filter set, $3<z<6$}\\
S/N $>$ 10 in-any filter & 1.410 & 0.062 & 0.480 & 33\\
m$_5<$ 26.6              & 1.273 & 0.036 & 0.311 & 13\\
m$_5<$ 25.6              & 0.455 & 0.026 & 0.040 & 2\\
m$_5<$ 24.6              & 0.036 & 0.024 & 0.006 & 0.2\\
\multicolumn{5}{c}{U-band filter set using D95 method, $0<z<3$}\\
S/N $>$ 10 in-any, D95$<$0.40 & 0.049 & 0.045 & 0.0015 & 103  \\
S/N $>$ 10 in-any, D95$<$0.30 & 0.042 & 0.039 & 0.0005 & 93  \\
S/N $>$ 10 in-any, D95$<$0.25 & 0.039 & 0.036 & 0.0004 & 86  \\
m$_5<$ 26.6, D95$<$0.40       & 0.046 & 0.043 & 0.0012 & 95  \\
m$_5<$ 26.6, D95$<$0.30       & 0.041 & 0.039 & 0.0005 & 89  \\
m$_5<$ 26.6, D95$<$0.25       & 0.039 & 0.036 & 0.0004 & 83\\
\hline
\multicolumn{5}{c}{17-band filter set, $0<z<3$}\\
S/N $>$ 10 in-any filter & 0.153 & 0.061 & 0.0214 & 133\\
m$_5<$ 26.6              & 0.144 & 0.052 & 0.0167 & 110\\
m$_5<$ 25.6              & 0.052 & 0.033 & 0.0010 & 67\\
m$_5<$ 24.6              & 0.025 & 0.025 & 0.0001 & 36\\
m$_5<$ 25.6, Early       & 0.024 & 0.024 & 0.0000 & 6\\
m$_5<$ 25.6, Late        & 0.039 & 0.037 & 0.0007 & 37\\
m$_5<$ 25.6, Starburst   & 0.071 & 0.028 & 0.0019 & 24
\enddata
\tablecomments{$\sigma_z\equiv rms[(z_{spec}-z_{phot})/(1 + z_{spec})]$. Outliers are defined as objects with
$|(z_{spec}-z_{phot})/(1 + z_{spec})|>0.3$. See \S5.1 for the definition of D95.
}
\label{t3}
\end{deluxetable}

\end{document}